\documentclass[12pt]{article}
\usepackage{latexsym,amssymb}
\usepackage[american]{babel}
\def\II{{\mathbb I}}
\def\RR{{\mathbb R}}

\def\tr{{\rm \,tr\,}}

\def\det{{\rm \,det\,}}

\def\be{\begin{equation}}
\def\ee{\end{equation}}
\def\bea{\begin{eqnarray}}
\def\eea{\end{eqnarray}}

\def\sideremark#1{\ifvmode\leavevmode\fi\vadjust{\vbox to0pt
{\vss\hbox to 0pt{\hskip\hsize\hskip1em
\vbox{\hsize2cm\tiny\raggedright\pretolerance10000
\noindent #1\hfill}\hss}\vbox to8pt{\vfil}\vss}}}

\begin{document}

\begin{titlepage}
\null

\centerline{\LARGE\bf A Noncommutative Deformation} 
\medskip 
\centerline{\LARGE\bf of General Relativity} 
\bigskip 
\centerline{\Large\bf Ivan G. Avramidi} 
\medskip 
\centerline{\it Department of Mathematics} 
\centerline{\it New Mexico Institute of Mining and Technology} 
\centerline{\it Leroy Place 801, Socorro, NM 87801, USA} 
\centerline{\it E-mail: iavramid@nmt.edu} 
\bigskip 

\begin{abstract}
We develop a novel approach to gravity in which 
gravity is described by a matrix-valued symmetric two-tensor
field and construct an  invariant functional 
that reduces to the standard Einstein-Hilbert action in the
commutative limit.  We also introduce a gauge symmetry
associated with the new degrees of freedom. 
\end{abstract}

\\ 

\end{titlepage}


It is expected that the general relativistic  description of gravity,
and, as the result, of the space-time, is inadequate at short distances.  
There are many different proposals   how to modify general relativity 
\cite{polchinski,konechny,rovelli}. An appealing idea is that  all these
approaches will be somehow related within one big unifying picture.

In this Letter we propose to describe gravity  by a matrix valued tensor
field  with a new gauge symmetry incorporated in the model. Our approach
should be contrasted with the ``noncommutative extensions of gravity''
on non-commutative spaces
\cite{chamseddine,chamseddine2,chamseddine3,madore}; it is also
different from the model studied in \cite{wald}.


The basic notions of general relativity are based on the geometrical
interpretation of the  wave equation that describe propagation of fields
{\it without internal structure}, (in particular, light). However, at
the microscopic distances the role of light (photon) could be played by
other gauge fields that, together with the photon, form a multiplet of
gauge fields {\it with some internal structure}. That is why 
one has to consider a linear wave equation for
such fields, i.e. instead of the scalar wave equation we have a {\it system}
of linear second-order hyperbolic (wave) partial differential equations.
This would cardinally change the standard geometric interpretation of
general relativity.  Exactly in the same way as a scalar equation
defines Riemannian geometry, a system of wave equations  will generate a
more general picture, that we call {\it Matrix Riemannian Geometry}.


Let us consider a multiplet of fields $\varphi=\left(\varphi^A(x)\right)$. 
The capital Latin indices 
denote the internal indices and run from $1$ to $N$.
We consider a {\it hyperbolic} system of linear second-order 
partial differential equations of the form
\be
L\varphi=
\left[a^{\mu\nu}(x)\partial_\mu\partial_\nu+b^\mu(x)\partial_\mu+c(x)\right]
\varphi
=0\,,
\ee
with {\it matrix-valued} coefficients
$a^{\mu\nu}=\left(a^{\mu\nu}{}^A{}_B(x)\right)$,
$b^{\mu}=\left(b^{\mu}{}^A{}_B(x)\right)$ and $c=\left(c^A{}_B(x)\right)$.
The matrix $a^{\mu\nu}$ is supposed to be Hermitian and 
symmetric in the vector indices, i.e.
\be
a^{\mu\nu}=a^{\nu\mu}\,, \qquad
\left(a{}^{\mu\nu}\right)^\dagger=a^{\mu\nu}\,.
\label{asymm}
\ee
Note that the matrix $a$ transforms under diffeomorphisms as a contravariant
matrix-valued two-tensor
\be
a'^{\mu\nu}(x')={\partial x'^{\mu}\over\partial x^\alpha}
{\partial x'^{\nu}\over\partial x^\beta}a^{\alpha\beta}(x)\,,
\label{diff}
\ee
We can also consider the
gauge transformations
\be
\varphi(x)\longrightarrow U(x)\varphi(x)\,,\qquad
a^{\mu\nu}(x)\longrightarrow U(x) a^{\mu\nu}(x)U(x)^{-1}\,,
\label{gt0}
\ee
where $U(x)$ is a nondegenerate matrix-valued function.

Let us consider the matrix
\be
H(x,\xi)=a^{\mu\nu}(x)\xi_\mu\xi_\nu\,,
\ee
where $\xi$ is a cotangent vector.
Obviously, it is Hermitian, $H^*=H$,
and, therefore, has real eigenvalues $h_i(x,\xi)$ 
(we will assume that their multiplicities $d_i$ are constant).
Further, the eigenvalues are invariant under the gauge transformations
(\ref{gt0}) and transform under the diffeomorphisms as
$h'_i(x',\xi)=h_i(x,\xi')$,
where 
$\displaystyle
\xi'_{\mu}={\partial x^\alpha\over\partial x'^\mu} \xi_\alpha$.

The system of hyperbolic partial differential equation describes the 
propagation of a {\it collection of waves}. 
The operator $L$ generates the causal structure on the manifold $M$ as
follows. First, we define the characteristics of the matrix hyperbolic
operator $L$. Each eigenvalue defines a Hamiltonian system, 
i.e. a Hamilton-Jacobi equation
\be
h_i(x,\partial S)=0\,,
\ee
and Hamilton equations
\be
{dx^\mu\over dt}={\partial\over\partial \xi_\mu} h_i(x,\xi)\,,
\qquad
{d \xi_\mu\over dt}=-{\partial \over\partial x^\mu} h_i(x,\xi)\,.
\label{hmm}
\ee
These equations define different trajectories for each eigenvalue.
The trajectories with tangents on the surface 
$h_i(x_0,\xi_0)=0$ are the null trajectories and define different causal cones
${\cal C}_i(x_0)$.
Each cone defines a causal set 
${\cal I}_i(x_0)$ consisting of the absolute past ${\cal I}_i^-(x_0)$
and the absolute future ${\cal I}_i^+(x_0)$ as well as the exterior of
the cone ${\cal E}_i(x_0)$. Since the points in all causal sets are
causally connected with the point $x_0$, i.e. there is at least one
time-like trajectory connecting those points with $x_0$, we define the
{\it causal set} as the union of all causal sets
$
{\cal I}(x_0)=\bigcup_{i=1}^s {\cal I}_i(x_0)\,,
$
similarly for the absolute past and the absolute future
$
{\cal I}^{\pm}(x_0)=\bigcup_{i=1}^s {\cal I}^{\pm}_i(x_0)\,,
$
The {\it causally disconnected set} is defined as the intersection
of the exteriors of all causal cones
$
{\cal E}(x_0)=\bigcap_{i=1}^s {\cal E}_i(x_0)\,.
$
With these definition we have the standard causal decomposition
$M={\cal I}^-(x_0)\cup{\cal I}^+(x_0)
\cup\partial {\cal I}(x_0)\cup{\cal E}(x_0)$.
Since the causal cones vary from point to point, the structure of the
causal set is different  at different points. Such a picture can be
interpreted as a ``{\it fuzzy light-cone}.''

We see that in the matrix case the operator $L$ does not define a unique
Riemannian metric. Rather there is a matrix-valued symmetric
2-tensor field $a^{\mu\nu}$.
Gravity will be described by new dynamical variable $a^{\mu\nu}$. Our
goal is to construct a diffeomorphism-invariant  functional $S(a)$
from the matrix $a^{\mu\nu}$ and its first derivatives, which
should be also invariant under gauge
transformations (\ref{gt0}) and reduce to the standard
Einstein-Hilbert functional in the commutative limit.

First of all, we need a measure, i.e. a density 
$\mu(a)$ that does not depend on the derivatives of $a$ and 
transforms under diffeomorphisms like $\mu'(x')=\mu(x)J(x)$,
where 
$\displaystyle
J(x)=\det\left[{\partial x'^{\mu}(x)\over \partial x^{\alpha}}\right]$.
We can write the measure in the form
\be
\mu={1\over N}\tr_V\rho\,,
\ee
where $\rho$ is a matrix-valued density, i.e. $\rho'(x')=\rho(x)J(x)$.
For example, $\rho$ can be defined by
\be
\rho(x)=\int\limits_{\RR^n}\, d\xi\, \Phi(x,\xi)\,,
\label{mames1}
\ee
where
$\Phi$ is a positive function of single variable such that it decreases
sufficiently fast as $\xi\to\infty$. 
The choice of the function
$\Phi$ should guarantee the convergence of this integral.
Note that this choice is obviously not unique. 
Another good candidate for the measure can be obtained as follows.
Let
\be
\psi={1\over n!}\varepsilon_{\mu_1\dots\mu_n}\varepsilon_{\nu_1\dots\nu_n}
a^{\mu_1\nu_1}\cdots  a^{\mu_n\nu_n}\,,
\ee
where $\varepsilon$ is the standard completely antisymmetric 
Levi-Civita symbol. 
The matrix $\psi$ is obviously Hermitian, $\psi^\dagger=\psi$.
We will require that this matrix is nondegenerate. Then the matrix
$\psi^\dagger\psi$ is positive definite so that we can define $\rho$ by
\be
\rho=(\psi^\dagger\psi)^{-1/4}\,.
\label{mames2}
\ee
The matrix $\rho$ is Hermitian, $\rho^\dagger=\rho$, 
and positive definite, $\rho>0$, so
that the measure is positive $\mu>0$.

To construct an invariant functional we need objects constructed 
from the derivatives of the matrix $a$ that transform like tensors.
First of all, we define another matrix-valued tensor $b_{\mu\nu}
=(b_{\mu\nu}{}^A{}_B)$
by
\be
a^{\mu\nu}b_{\nu\lambda}
=b_{\mu\nu}a^{\nu\lambda}
=\delta^\mu_\lambda\II\,.
\label{andg}
\ee
One can easily show that the matrix $b_{\mu\nu}$ satisfies the equation
\be
b^\dagger_{\mu\nu}=b_{\nu\mu}\,,
\ee
but is not necessarily a Hermitian matrix 
symmetric in tensor indices, more precisely,
$b^\dagger_{\mu\nu}\ne b_{\mu\nu}$, $b_{\mu\nu}\ne b_{\nu\mu}$.

Next we introduce  matrix-valued coefficients 
${\cal A}^\mu{}_{\alpha\beta}=
\left({\cal A}^\mu{}_{\alpha\beta}{}^A{}_B\right)$ that
transform like the connection coefficients under the diffeomorphisms, i.e.
\be
{\cal A}'^{\mu'}{}_{\alpha'\beta'}(x')={\partial x'^{\mu}\over\partial x^\nu}
{\partial x^{\gamma}\over\partial x'^{\alpha}}
{\partial x^{\delta}\over\partial x'^{\beta}}
{\cal A}^\nu{}_{\gamma\delta}(x)
+{\partial x'^{\mu}\over\partial x^{\nu}}
{\partial^2 x^{\nu}\over\partial x'^\alpha\partial x'^\beta}
\II\,.
\ee
We use these "matrix-valued connection" to define an
operator ${\cal D}$ acting on tensors of type $(p,q)$
by
\bea
{\cal D}_\alpha\varphi^{A\mu_1\dots\mu_p}_{\nu_1\dots\nu_q}
&=&\partial_\alpha\varphi^{A\mu_1\dots\mu_p}_{\nu_1\dots\nu_q}
+\sum_{j=1}^p{\cal A}^{\mu_j}{}_{\lambda\alpha}{}^A{}_B
\varphi^{B\mu_1\dots\mu_{j-1}\lambda\mu_{j+1}\dots\mu_p}_{\nu_1\dots\nu_q}\,
\nonumber\\
&&
-\sum_{i=1}^q {\cal A}^{\lambda}{}_{\nu_i\alpha}{}^A{}_B
\varphi^{B\mu_1\dots\mu_p}_{\nu_1\dots\nu_{i-1}\lambda\nu_{i+1}\dots\nu_q}
\,.
\eea

Now it is not difficult to construct the "matrix-valued curvature" and the
"matrix-valued torsion" by
\be
({\cal D}_\mu{\cal D}_\nu
-{\cal D}_\nu{\cal D}_\mu)\varphi_{\alpha}
=-{\cal R}^{\lambda}{}_{\alpha\mu\nu}\varphi_\lambda
+{\cal T}^\lambda{}_{\mu\nu}{\cal D}_\lambda\varphi_{\alpha}\,,
\ee
where
\bea
{\cal R}^\lambda{}_{\alpha\mu\nu}
&=&\partial_\mu {\cal A}^\lambda{}_{\alpha\nu}
-\partial_\nu {\cal A}^\lambda{}_{\alpha\mu}
+{\cal A}^\lambda{}_{\beta\mu}{\cal A}^\beta{}_{\alpha\nu}
-{\cal A}^\lambda{}_{\beta\nu}{\cal A}^\beta{}_{\alpha\mu}\,,
\\[10pt]
{\cal T}^\lambda{}_{\mu\nu}
&=&{\cal A}^\lambda{}_{\mu\nu}-{\cal A}^\lambda{}_{\nu\mu}\,.
\eea

To fix the connection we impose the {\it compatibility
condition}. Unfortunately, such a condition is {\it not unique}.
We take it in the form 
\be
\partial_\mu a^{\alpha\beta}
+{\cal A}^\alpha{}_{\lambda\mu}a^{\lambda\beta}
+{\cal A}^\beta{}_{\lambda\mu}a^{\alpha\lambda}=0\,
\label{19}
\ee
(which enables one to find an exact explicit
solution of ${\cal A}^\alpha{}_{\beta\gamma}$ in terms of 
derivatives of the matrix $a^{\mu\nu}$)
and require that the connection must be 
{\it symmetric in the commutative limit},
i.e. the torsion is purely a deformation artifact.
The solution of these constraints is
\bea
{\cal A}^\alpha{}_{\lambda\mu}
&=&{1\over 2}b_{\lambda\sigma}
\Biggl(
a^{\alpha\gamma}\partial_\gamma a^{\rho\sigma}
-a^{\rho\gamma}\partial_\gamma a^{\sigma\alpha}
-a^{\sigma\gamma}\partial_\gamma a^{\alpha\rho}
\Biggr)b_{\rho\mu}\,,
\eea


By using the matrix curvature
we can now construct a simple
generalization of the standard Einstein-Hilbert functional
(with cosmological constant). We define
\be
S_0(a)=\int\,dx\,{1\over 16\pi G}{1\over N}\tr_V 
\rho\left(a^{\mu\nu}{\cal R}^\alpha{}_{\mu\alpha\nu}-2\Lambda\right)\,.
\ee
This functional is obviously invariant under global gauge transformations.

One can easily make it local gauge symmetry by introducing a Yang-Mills
field ${\cal B}$
and replacing the partial derivatives in the definition
of the connection coefficients and the curvature by covariant derivatives.
Of course, this must be also done in the compatibility condition
(\ref{19}) for that condition to be consistent with the gauge invariance.

%
%
%
%
Thus we obtain finally a gauged version of the above functional
\bea
S(a,{\cal B})&=&\int\limits_\Omega\,dx\,
{1\over N}\tr_V \rho\Biggl\{{1\over 16\pi G}
\left(a^{\nu\mu}{\cal R}^\alpha{}_{\mu\alpha\nu}-2\Lambda\right)
-{1\over 2e^2}
a^{\nu\mu}{\cal F}_{\mu\alpha}a^{\alpha\beta}{\cal F}_{\beta\nu}
\Biggr\}\,,
\eea
where $e$ is the Yang-Mills coupling constant,
$\rho$ is defined by eq. (\ref{mames1}) or (\ref{mames2})
\bea
{\cal R}^\lambda{}_{\alpha\mu\nu}
&=&\partial_\mu {\cal A}^\lambda{}_{\alpha\nu}
+[{\cal B}_\mu,{\cal A}^\lambda{}_{\alpha\nu}]
-\partial_\nu {\cal A}^\lambda{}_{\alpha\mu}
-[{\cal B}_\nu, {\cal A}^\lambda{}_{\alpha\mu}]
\nonumber\\[10pt]
&&
+{\cal A}^\lambda{}_{\beta\mu}{\cal A}^\beta{}_{\alpha\nu}
-{\cal A}^\lambda{}_{\beta\nu}{\cal A}^\beta{}_{\alpha\mu}
\eea
\bea
{\cal A}^\alpha{}_{\lambda\mu}
&=&b_{\lambda\sigma}
\Biggl\{
{1\over 2}\Bigl[
a^{\alpha\gamma}\partial_\gamma a^{\rho\sigma}
+a^{\alpha\gamma}[{\cal B}_\gamma, a^{\rho\sigma}]
\nonumber\\
&&
-a^{\rho\gamma}\partial_\gamma a^{\sigma\alpha}
-a^{\rho\gamma}[{\cal B}_\gamma, a^{\sigma\alpha}]
-a^{\sigma\gamma}\partial_\gamma a^{\alpha\rho}
-a^{\sigma\gamma}[{\cal B}_\gamma, a^{\alpha\rho}]
\Bigr]
\Biggr\}b_{\rho\mu}\,,
\label{24}
\eea
\be
{\cal F}_{\mu\nu}
=\partial_\mu {\cal B}_{\nu}
-\partial_\nu {\cal B}_{\mu}
+[{\cal B}_\mu,{\cal B}_\nu]\,.
\ee
Notice that (\ref{24}) is nothing but the solution of the gauged
version of the compatibility condition
\be
\partial_\mu a^{\alpha\beta}
+[{\cal B}_\mu,a^{\alpha\beta}]
+{\cal A}^\alpha{}_{\lambda\mu}a^{\lambda\beta}
+{\cal A}^\beta{}_{\lambda\mu}a^{\alpha\lambda}=0\,.
\label{26}
\ee
Here, according to our construction,
the matrix affine connection ${\cal A}^\alpha{}_{\beta\gamma}$
acts only from the left, however, the gauge connection ${\cal B}_\mu$
acts from {\it both} left and right as any gauge connection should.

Now we need to introduce the spontaneous
breakdown of the gauge symmetry, so that in the broken phase in the
vacuum there is just one tensor field, which is identified with the
metric of the space-time. All other tensor fields must have zero vacuum
expectation values. In the unbroken phase there will not be a metric at
all in the usual sense since there is no preferred tensor field with
non-zero vacuum expectation value. Alternatively, one could expect the
gauge ({\it gravicolor}) degrees of freedom to be confined within the Planck
scales, so that only the invariants (graviwhite states) are visible at
large distances. For example, at large distances one could only see the
diagonal part $g^{\mu\nu}={1\over N}\tr_V a^{\mu\nu}$, which defines the
metric of the space-time at large distances and the gauge invariants
like $ {1\over N}\tr_V{\cal R}^\mu{}_{\alpha\beta\gamma}\,, $ etc.,
which determine in some sense the curvature of the spacetime.

This model may be viewed as a ``noncommutative 
deformation'' of Einstein gravity coupled to a Yang-Mills model.
In the weak deformation limit our
model describes Einstein gravity, Yang-Mills fields, and a multiplet of 
self-interacting two-tensor fields that interact also with gravity and
the Yang-Mills fields. We speculate that the new degrees of freedom
could only be visible at Planckian scales, so that they do not exhibit
themselves in the low-energy physics. However, the behavior of our model
at higher energies should be radically different from the Einstein
gravity since there is no preferred metric in the unbroken phase, when
the new gauge symmetry is intact. 

Our model is, in fact, nothing but a generalized sigma
model. So, the problems in quantization of this model are the same as in the
quantization of the sigma model. We point out that the
study of the one-loop approximation requires 
new calculational 
methods since the partial differential operators involved are not
of the so-called Laplace type (nonscalar leading symbol).
Most of the calculations in quantum field theory were restricted so far to the
Laplace type operators for which nice theory of heat kernel asymptotics is
available. However, the study of heat kernel asymptotics for non-Laplace type
operators is quite new and the methodology is still underdeveloped. For example,
even the first heat kernel coefficients ($A_0$, $A_1$ and $A_2$) 
needed for the renormalization in four dimensions are not known in general. 
For some progress in this area (the calculation of $A_1$) see
\cite{avrbran01,avrbran02}.

\bigskip

I am grateful to Giampiero Esposito for stimulating discussions. The
financial support by the University of Naples and the
Istituto Nazionale di Fisica Nucleare is gratefully acknowledged.


\begin{thebibliography}{999}

\bibitem{polchinski} J. Polchinski, {\it String Theory},
(Cambridge: Cambridge University Press, 1999)

\bibitem{konechny}
A. Konechny and A. Schwarz, {\it 
Introduction to M(atrix) theory and noncommutative geometry},
Phys. Repts.  {\bf 360} (2002) 353--465

\bibitem{rovelli}
C. Rovelli, {\it Loop Quantum Gravity}, Living Rev. Rel. {\bf 1} (1998) 1,
gr-qc/9710008 

\bibitem{chamseddine}
A. H. Chamseddine, G. Felder, J. Fr\"ohlich,
{\it Gravity in Non-Commutative Geometry},
Commun. Math. Phys. 155 (1993) 205--218

\bibitem{chamseddine2}
A. H. Chamseddine, {\it Noncommutativity Gravity},
hep-th/0301112

\bibitem{chamseddine3}
A. H. Chamseddine,
{\it An Invariant Action for Noncommutative Gravity in Four-Dimensions},
J. Math. Phys. {\bf 44} (2003) 2534-2541

\bibitem{madore} 
J. Madore and J. Mourad, {\it A noncommutative extension 
of gravity}, Int. J. Mod. Phys. D {\bf 3} (1994) 221--224

\bibitem{wald} 
R. M. Wald, {\it A new type of gauge invariance for a collection
of massless spin-2 fields: II. Geometrical interpretation},
Class. Quantum Grav. {\bf 4} (1987) 1279--1316

\bibitem{avrbran01} 
I. G. Avramidi and T. Branson, 
{\it Heat kernel asymptotics of operators with non-Laplace principal part}, 
Rev. Math. Phys. {\bf 13} (2001) 847--890

\bibitem{avrbran02} 
I. G. Avramidi and T. Branson, 
{\it A discrete leading symbol and spectral asymptotics for natural differential 
operators}, J. Funct. Anal. {\bf 190} (2002) 292--337  

\end{thebibliography}
\end{document}